\documentclass[conference]{IEEEtran}

\usepackage{graphicx}
\graphicspath{ {/} }

\usepackage[nocompress]{cite}

\usepackage{hyperref}
\usepackage{enumerate}
\usepackage{color}

\usepackage{tabulary}
\usepackage{booktabs}
\usepackage{makecell}
\usepackage{multirow}
\usepackage{setspace}
\usepackage{array}

\usepackage{todonotes}

\usepackage{amssymb}

\usepackage{amsmath}

\renewcommand{\arraystretch}{1.1}
\newcommand{\nlcell}[3][1.0]{\def\arraystretch{#1}\begin{tabular}[c]{@{}#2@{}}#3\end{tabular}}

\begin{document}

\title{Customer Support Ticket Escalation Prediction using Feature Engineering}
% \subtitle{Do you have a subtitle?\\ If so, write it here}

%\titlerunning{Short form of title}        % if too long for running head
% \author{
%     Author 1, Author 2, Author 3
%     \\ University
%     \\ Location 1
%     \\ \{author1, author2, author3\}@university.tld
%     \and
%     Author 4
%     \\ Association, Company
%     \\ Location 2
%     \\ author4@company.tld
% }
\author{
    Lloyd Montgomery, Daniela Damian, Tyson Bulmer
    \\ University of Victoria
    \\ Victoria, Canada
    \\ \{lloydrm, danielad, tysonbul\}@uvic.ca
    \and
    Shaikh Quader
    \\ Private Cloud Platform Digital Support, IBM
    \\ Toronto, Canada
    \\ shaikhq@ca.ibm.com
}

% UNCOMMENT FOR REJ FORMAT
% \institute{}
% \institute{
%     Lloyd Montgomery, Daniela Damian, Tyson Bulmer
%     \at University of Victoria \\
%     \email{\{lloydrm, danielad, tysonbul\}@uvic.ca}
%         \and
%     Shaikh Quader
%     \at IBM Canada, Private Cloud Platform Digital Support \\
%     \email{shaikhq@ca.ibm.com}
% }

\date{Received: date / Accepted: date}
% The correct dates will be entered by the editor

\maketitle

% Main Content of Journal Article

\begin{abstract}

Understanding and keeping the customer happy is a central tenet of requirements engineering. Strategies to gather, analyze, and negotiate requirements are complemented by efforts to manage customer input after products have been deployed. For the latter, support tickets are key in allowing customers to submit their issues, bug reports, and feature requests. If insufficient attention is given to support issues, however, their escalation to management becomes time-consuming and expensive, especially for large organizations managing hundreds of customers and thousands of support tickets. Our work provides a step towards simplifying the job of support analysts and managers, particularly in predicting the risk of escalating support tickets. In a field study at our large industrial partner, IBM, we used a design science research methodology to characterize the support process and data available to IBM analysts in managing escalations. In a design science methodology, we used feature engineering to translate our understanding of support analysts' expert knowledge of their customers into features of a support ticket model. We then implemented these features into a machine learning model to predict support ticket escalations. We trained and evaluated our machine learning model on over 2.5 million support tickets and 10,000 escalations, obtaining a recall of 87.36\% and an 88.23\% reduction in the workload for support analysts looking to identify support tickets at risk of escalation. Further on-site evaluations, through a prototype tool we developed to implement our machine learning techniques in practice, showed more efficient weekly support-ticket-management meetings. Finally, in addition to these research evaluation activities, we compared the performance of our support ticket model with that of a model developed with no feature engineering; the support ticket model features outperformed the non-engineered model. The artifacts created in this research are designed to serve as a starting place for organizations interested in predicting support ticket escalations, and for future researchers to build on to advance research in escalation prediction.

\end{abstract}

\section*{Acknowledgements}

We thank IBM for their data, advice, and time spent as a collaborator: Keith Mackenzie at IBM Victoria for his time and access to his team, Jean-Francois Puget for his guidance towards using XGBoost, and Chad Marston and Tracy Bolot for management and funding support. We thank Emma Reading for her contribution to the prototype tool. We thank the anonymous referees of both RE17 and the REJ special issue. This research was funded by the Natural Sciences and Engineering Research Council of Canada (NSERC) and IBM Center for Advanced Studies (IBM CAS). This is a pre-print of an article published in Springer Requirements Engineering Journal. The final authenticated version is available online at: \url{https://doi.org/10.1007/s00766-018-0297-y}.

\section{Introduction}

Large software organizations handle many customer support issues every day in the form of bug reports, feature requests, and general misunderstandings as submitted by customers. A significant portion of these issues create new, or relate to, existing technical requirements for product developers, thus allowing requirements management and release planning processes to be reactive to customer input.

These support issues are submitted through various channels such as support forums and product wikis, however, a common default for organizations is to offer direct support through phone and online systems in which support tickets are created and managed by support analysts. The process of addressing these support tickets varies across different organizations, but all of them share a common goal: to resolve the issue brought forth by the customer and keep the customer happy. If a customer is not happy with the support they are receiving, companies have escalation processes whereby customers can state their concern for how their support ticket is being handled by escalating their problems to management's attention.

While the escalation process is needed to draw attention to important and unresolved issues, handling the underlying support ticket after an escalation occurs is very expensive for organizations \cite{Ling2005}, amounting to millions of dollars each year \cite{sheng2014cost}. Additionally, gathering bottom-up requirements from support tickets is an important requirements-gathering practice for companies looking to address customer feedback and suggestions; however, escalations (and the process of managing them) take time away from support analysts, making the discovery of bottom-up requirements much less efficient. When escalations occur, immediate management and senior software engineers' involvement is necessary to reduce the business and financial loss to the customer. Furthermore, software defect escalations can --- if not handled properly --- result in a loss of reputation, satisfaction, loyalty, and customers \cite{Boehm1984}.

Understanding the customer is a key factor in keeping them happy and solving support issues. It is the customer who, driven by a perceived ineffective resolution of their issue, escalates tickets to management's attention \cite{Bruckhaus2004}. A support analyst's job is to assess the risk of support-ticket escalation given the information present --- a largely manual process. This information includes the customer, the issue, and interrelated factors such as time of year. Keeping track of customers and their issues becomes infeasible in large organizations who service multiple products across multiple product teams, amounting to large amounts of customer data.

Past research proposed Machine Learning (ML) techniques that model industrial data and predict escalations \cite{Ling2005, sheng2014cost, Bruckhaus2004, Marcu2009}, though none of these efforts attempted to equip ML algorithms with the knowledge-gathering techniques that support analysts use every day to understand their customers. The focus had instead been on improving Escalation Prediction (EP) algorithms while utilizing largely all available support data in the studied organization, without much regard to modelling analysts' understanding of whether customers might escalate. Defining which information analysts use to identify issues at risk of escalation is the first step in Feature Engineering (FE): a difficult, expensive, domain-specific task of finding features that correlate with the target class \cite{Domingos2012} (in this case, escalations). In our research we conducted FE to describe customer escalations, driven by the following research question:

\begin{description}
    \item[RQ 1.] What are the features of a support-ticket model to best describe a customer escalation?
\end{description}

The ``support ticket model" is a set features engineered to capture elements of the support ticket and escalation process so that, when data is mapped to those features and fed into a ML model, the process of predicting escalations is improved (when compared to an approach with no feature engineering). Since these features leverage the context around the analysts' work, we then explored the use of these features within ML models in our efforts to automate some parts of the analysts' EP and management process: 

\begin{description}
    \item[RQ 2.] Can ML techniques that implement such a model assist in escalation management?
\end{description}

Finally, acknowledging that FE is a task that requires both time to conduct and knowledge of the underlying contextual system that is trying to be modelled, we sought to evaluate the performance of the ML models leveraging FE efforts in our research against ML models that don't harness FE efforts.

\begin{description}
    \item[RQ 3.] Does FE improve ML results over using all available customer support ticket data?
\end{description}

In answering our research questions, the contributions of our work have been iteratively developed and evaluated through a Design Science Research Methodology \cite{VonAlan2004, Sedlmair2012, Wieringa2014} in collaboration with our industrial partner, IBM. Our first main contribution is the Support Ticket Model Features –-- through FE –-- that support teams use to assess and manage the risk of escalations. This contribution was developed through observations of practice and interviews with management, developers, and support analysts at IBM, as well as analysis of the IBM customer support data repository. Our second contribution is the investigation of this model when used with ML techniques to assist in the escalation process. We complemented a statistical validation of our techniques with an in-depth study of the use of these techniques in daily management meetings assessing escalations at one collaborating product team, IBM Victoria in Canada. Finally, we show that FE added value to the ML results by implementing a baseline in which no FE was conducted, and comparing the performance of the models we developed with and without FE.

The work reported here was originally published and presented at the \emph{25th International Conference on Requirements Engineering (RE'17)} \cite{Montgomery2017Full}. The conference paper reported on the first two evaluation cycles in our design science methodology. This article revises the RE'17 paper and extends it in several ways:

\begin{itemize}
    \item We engineered additional features in our Support Ticket Model to incorporate feedback from the first two evaluation cycles in our methodology. This required further processing of our data and resulted in a more complete final set of features in our model.
    \item We added a third evaluation cycle to our design science methodology to validate, through statistical methods, the performance of the final model including all features developed through this research study. This evaluation cycle also involved switching to a new algorithm, XGBoost, to improve the precision of our model results.
    \item We added a new research question to investigate the gain in model performance from of our FE efforts. A fourth evaluation cycle in our design science methodology was added to develop a baseline model with no FE efforts and to compare its performance to that of the models we developed through FE.
\end{itemize}

\section{Related Work}

The development and maintenance of software products is highly coupled with many stakeholders, among which the customer plays a key role. Software Product Management (SPM) is a large area of research that covers many facets of software products. As proposed by van de Weerd et al. \cite{VandeWeerd2006}, within SPM is portfolio management, product roadmapping, requirements management, and release planning. Our research is concerned with providing support for a product, which in the above categories comes out as consequence of release planning, and then feeds back into requirements management through (bottom up) requirements gathering. However, the broader category of which this research fits into is Customer Relationship Management (CRM), which involves integrating artifacts, tools, and workflows to successfully initiate, maintain, and (if necessary) terminate customer relationships \cite{Reinartz2013}. Although all of the above categories of SPM involve some amount of CRM, CRM is a subset of SPM. Examples of CRM practices include: customer participation requirements-gathering sessions, customer feature suggestions through majority voting, customer incident reports, and support tickets \cite{Kabbedijk2009, Merten2016}. Other tactics of involving customers in the requirements gathering phase such as stakeholder crowd-sourcing (e.g. Lim et al. \cite{Lim2011}) and direct customer participation (e.g. Kabbedijk et al. \cite{Kabbedijk2009}) are CRM processes that aim to mitigate the potential cost of changing-requirements after development has begun.

An outstanding aspect, however, is the effort and cost associated with the management of a product's ongoing support process: dealing with bugs, defects, and feature requests through processes such as product wikis, support chat lines, and support tickets. When support tickets are not handled in a timely manner or a customer's business is seriously impacted, customers escalate their issues to management \cite{sheng2014cost}. Escalation is a process very costly for organizations \cite{sheng2014cost, Bruckhaus2004} and yet fruitful for research in ML that can parse large amounts of support ticket data and suggest escalation trends \cite{Bruckhaus2004, Berry1997}.

ML techniques have been proposed in various ways in previous research to address EP. Marcu et al. \cite{Marcu2009} used a three-stage correlation and filter process to match new support issues with existing issues in the system. Their goal and contribution was to speed up the triage and resolution process through finding similar issues previously resolved. Ling et al. \cite{Ling2005} and Sheng et al. \cite{sheng2014cost} propose cost-sensitive learning as a technique for improved ML results optimized for EP. Their research, however, was primarily focused on the cost-sensitive learning algorithms and the improvements they offered, with no consideration to the individual features being fed into the model. Similarly, Bruckhaus et al. \cite{Bruckhaus2004} conducted preliminary work investigating the use of neural networks to conduct EP on data from Sun Microsystems. Their work does not describe how they selected their final features from an initial set of 200.

A similar field to EP is bug prediction, where research reports significant efforts in ML techniques for bug prediction. Although similar in prediction efforts, the target outcomes differ significantly in the two fields of research. EP is trying to predict escalations, which are outcomes driven mostly by customers, whereas bug prediction is trying to predict bugs and faults within software, which are outcomes driven mostly by the structure of the software itself. There is an argument to be made that perhaps the developers and their environment contribute to the bugs and faults introduced into the software, but that is outside the scope of both this paper and the related work of bug prediction discussed in this section. A notable similarity between EP and bug prediction is the categories of artifacts used to perform the predictions. Research into bug prediction is mostly split between two artifact types: change log analysis approaches, and single version analysis approaches \cite{DAmbros2010}.

Change log analysis approaches utilize historical data, attempting to learn from how data has changed over time. The type of data being used includes code repositories to analyze code churn \cite{Nagappan2005,Moser2008,Hassan2009}, and past bug and defect reports \cite{Hassan2005,Ostrand2005,Bernstein2007,Kim2007,Moser2008}. Our research also utilizes historical data, but we neither utilize code repositories nor do we utilize bug and defect reports directly. Due to the nature of customer support tickets, it is common for a support ticket to cause a bug report to be created in response to the customer's issue (if the issue involves a bug with the software); however, these are different types of artifacts containing different types of information.

Single version analysis approaches do not utilize historical data, rather, they focus on the latest version of artifacts. As stated by D'Ambros and Robbes\cite{DAmbros2010}, ``single-version approaches assume that the current design and behavior of the program influences the presence of future defects". Our research also utilizes the most recent version of artifacts to build some of the features presented in this paper. Past history plays a role in whether support tickets will escalate or not, and so does the current state of their support ticket.

The end goal of EP through ML is to identify events generated by customers which might lead to escalations, yet none of the previous research attempts to solve the problem of EP by understanding how analysts identify escalations. Previous research does not focus on the customer through data selection or FE aimed at the knowledge that support analysts have about their customers. Our work addresses this by doing several iterative phases: extensive context-building work within a support organization; iterative cycles of FE focused on understanding the analysts' knowledge of the customer during the support ticket and escalation management process; and finally, real-world deployment of our ML techniques that implement this model to gain feedback on the Support Ticket Model Features.

Finally, to guide and implement the iterative phases of research and implementation described above, we employed a design science methodology. Inspired by the work of Simon \cite{simon1981sciences}, March and Smith \cite{March1995} originally introduce design science as attempts to create things that serve human purposes. It later became a popular, accepted research methodology in information sciences, due to the highly cited guidelines developed by Hevner \cite{VonAlan2004}. Design science methodology enables the design and validation of solution proposals to practical problems, and, as Wieringa puts it in the context of software engineering research \cite{Wieringa2009}, design science research is based on a very close connection between artifact design and research. In Hevner et al.\cite{VonAlan2004} guidelines, (1) a organization's business needs drive the development of validated artifacts that meet those needs, and (2) the knowledge produced in the development of these artifacts can be added to the shared research knowledge base. In our work, we grounded our research in the application of FE and ML in the context of the escalation problem at IBM, and specifically the development and evaluation of solutions to this problem.

\section{Design Science Research Methodology}

This research began when IBM approached our research team because of our previous empirical work \cite{Schroter2012, Wolf2009} in investigating development practice in IBM software teams and developing ML solutions to support developer coordination. A large organization offering a wide range of products to many customers world-wide, IBM described their current problem as: an increasing number of customer issue escalations resulting in additional costly efforts, as well as dissatisfied customers. They sought some automated means to enhance their support process through leveraging the data available in their large customer support repository.

\begin{figure*}[ht]
    \centering
    \includegraphics[width=\textwidth]{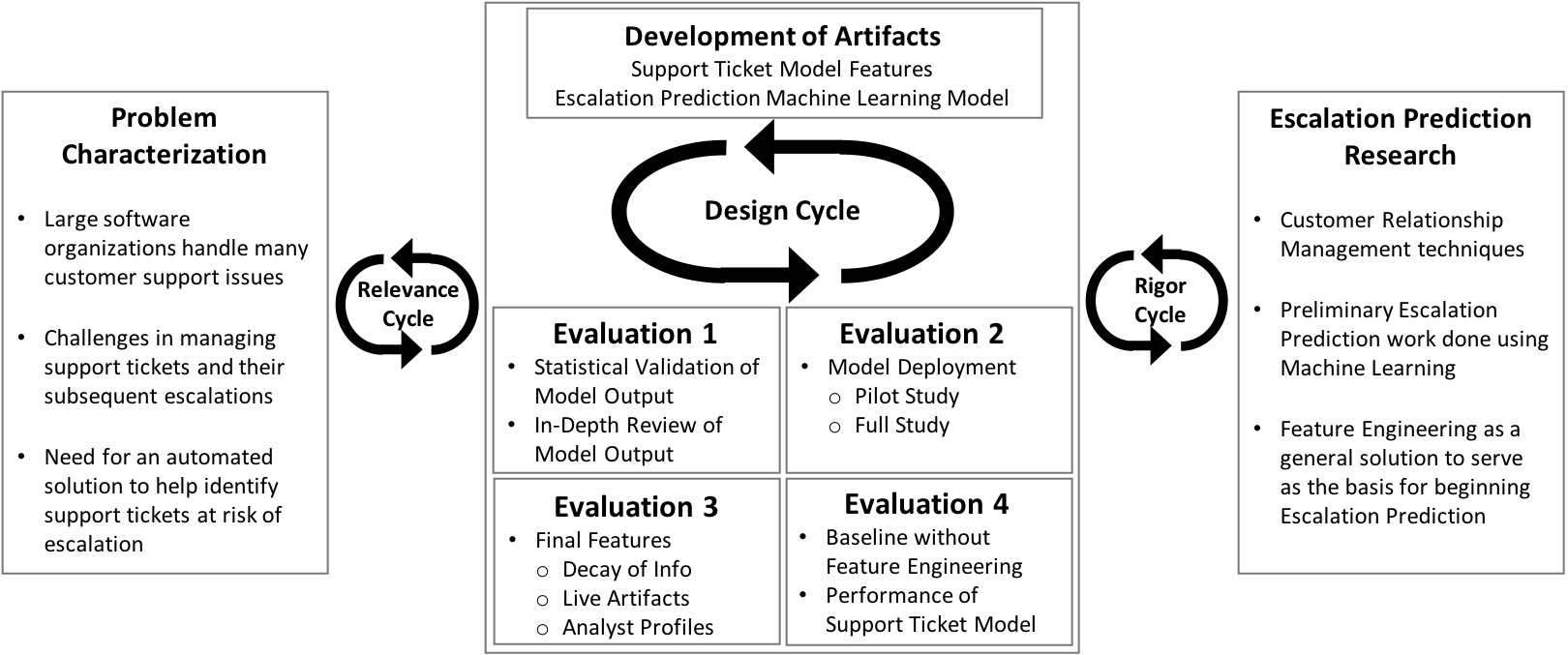}
    \caption{Design Science Research Methodology}
    \label{fig:methodology}
\end{figure*}

To investigate this problem and to develop techniques to support the analysts' job in the the escalation process, we employed a design science methodology \cite{VonAlan2004, Sedlmair2012, Wieringa2014}. As illustrated in Fig. \ref{fig:methodology}, our methodology iteratively developed and evaluated techniques to enhance the IBM support process from an understanding of the problem domain and close interaction with its stakeholders. Below we describe the steps and the process of our design science methodology in more detail.

\subsection{Problem Characterization}
We conducted an ethnographic exploratory study of the escalation process and data available to IBM customer support analysts. We interacted closely with the management and support team at the IBM Victoria site, which employs about 40 people working on two products called IBM Forms and Forms Experience Builder. Several other IBM employees in senior management, worldwide customer support, and Watson Analytics provided us with their input about the support process. Section \ref{sec:ProblemChar} details our ethnographic exploratory study and the insights about the problem in the IBM's escalation process as we came to understand it. 

\subsection{Research Artifact Development and Evaluation Stages}
We iterated through the development and evaluation of two artifacts in collaboration with our industrial partner: (1) the Support Ticket Model Features (RQ1) which represents the contextual knowledge held by support analysts about the support process, and (2) an EP ML model (RQ2) that represents the operationalization of the Support Ticket Model Features into a ML model to predict support ticket escalations. Section \ref{sec:stmf_features} outlines the Support Ticket Model Features as we developed them through the iterative cycles of our design science methodology. A first set of model features were developed through an ethnographic study at IBM during the Problem Characterization phase, as described in Section \ref{ssec:original_model}. This was followed by a few rounds of evaluations of our model, by means of developing and testing the performance of a ML model that implemented the Support Ticket Model Features to predict escalations (RQ2).

Evaluation 1 (Section \ref{sec:eval1}) involved the creation and statistical validation of a ML model that implemented this first set of features in our Support Ticket Model, as well as an in-depth review of the ML model output with IBM. The creation of the ML model involved feeding our support ticket data into multiple ML algorithms including CHAID, SVM, Logistic Regression, and Random Forest. Once the results could be analyzed across all of the implementations, the algorithm that produced the highest recall was selected. The in-depth review of the ML model output (Section \ref{ssec:model_output_validation}) was a 2-hour review session in which IBM managers, developers, and support analysts discussed the output of ten important support ticket escalations and compared their experience of the support ticket to the output of the model. This evaluation resulted in new and modified features into our Support Ticket Model. 

Evaluation 2 (Section \ref{sec:eval2}) used a Web implementation to deliver the results of the ML model to IBM to support analysts and management so they could utilize the results by integrating them into their work-flow. The tool was deployed for four weeks and used by support analysts and managers addressing support tickets. This evaluation resulted in new features into our Support Ticket Model. 

Evaluation 3 (Section \ref{sec:eval3}) was another round of statistical validation, and this time the model included the new features developed through Evaluations 1 and 2. This combined set of features (deemed the ``final features") was evaluated and compared to the first features through confusion matrices. Additionally, a new ML model, XGBoost, was implemented following feedback from our industrial partner, IBM. XGBoost produced much more diverse PR-Space graphs that gave us more options in selecting trade-offs in precision and recall that Random Forest did not.

The fourth and last evaluation (Section \ref{sec:eval4}) involved feeding the available support ticket data into the ML algorithm with as little manipulation as possible to validate that the FE efforts conducted were producing higher results than a model without any engineered features.

\subsection{Escalation Prediction Research}
Finally, to fulfill the rigor cycle in our methodology, we reviewed existing work in CRM and EP through ML, and reflected on how our research results are transferable to other settings.

In the remainder of the paper we describe in detail the Support Ticket Model Features as developed incrementally and iteratively through the rounds of empirical evaluations. Before then, however, we start by describing in Section \ref{sec:ProblemChar} the ethnographic exploratory study and its findings as part of our Problem Characterization phase. 

\section{Problem Characterization}
\label{sec:ProblemChar}

To ground the development of the two artifacts in a deeper understanding of the problem expressed by IBM, we first conducted an ethnographic exploratory study of the IBM support ticket process and escalation management practice. In this section, we discuss the details of our study and the insights we obtained towards a detailed characterization of the problem and its context.

\subsection{Ethnographic Exploratory Study and the IBM Escalation Process}
\label{ssec:ethno}

To learn about IBM processes, practices, and tools used by support analysts to collect and manage customer support tickets, one of the researchers worked on-site at IBM Victoria for two months. He attended daily support stand-up meetings run jointly by development and support management, and conducted follow-up interviews with management, developers and support analysts. The IBM Victoria staff involved in these sessions included the Victoria Site Manager, the Development Manager, the L3 Support Analyst, and two L2 Support Analysts. Additional information about the IBM support ticket process and escalation management practice was sought through interviews with four other senior analysts and managers at IBM support organizations in North Carolina and California. Additionally, extensive time was spent understanding the data available in the large IBM support ticket repository. We obtained customer support data consisting of 2.5 million support tickets and 10,000 escalation artifacts from interactions with 127,000 customers in 152 countries.

IBM has a standard process for recording and managing customer support issues across all its products. The support process involves multiple levels: L0, ownership verification; L1, basic user-error assistance; L2, product usage assistance from knowledge experts; and L3, development support of bugs and defects. 

\subsubsection{Support Level L0}
When a new support issue is filed by a customer, a Problem Management Record (PMR) is created by L0 to document the lifetime of the issue (for simplicity, we may use the term PMR to refer to a support ticket henceforth in the paper). The role of L0 is to verify that the customer owns the product they are seeking support for. If verified, the customer is then directed to L1 support. 
\subsubsection{Support Level L1}
L1 support is offered in the user's native language, by people who are qualified to help customers through basic support of most products offered by IBM. Due to the broad range of products that are supported by L1, they are not experts in any one product; therefore, if L1 is unable to solve the customer's problem --- or the problem is thought to be with the product itself (bug, usability, etc) --- the customer is then transferred to L2 support. \subsubsection{Support Level L2}
L2 support is offered by direct employees of the product the customer is seeking support on, so the customer is now dealing with an expert in the product they are seeking help for. Possible directions for L2 support analysts at this stage include one-on-one help walking through an issue with the customer, communicating with developers to get information on how the system should be behaving, and guidance from L3 support. 
\subsubsection{Support Level L3}
L3 support analysts are regarded as the most knowledgeable product experts for the product they support, and it is common for the L3 role to be filled by developers of the product they support, who rotate through the role. These support analysts are in charge of the more severe, nuanced, and time-consuming issues. Although PMRs can technically escalate at L1 and above, they normally escalate while being handled by L2 or L3 support analysts. 

IBM handles escalations through a process, and artifact, called a Critical Situation (CritSit) that is used when customers are not happy with the progress of their PMR. A PMR is said to ``Crit" when a CritSit is opened and that PMR is attached to the CritSit artifact. CritSits can be opened by customers for any reason, although the most likely scenario is to speed up the resolution of their PMR for business or financial reasons. The process of opening and handling a CritSit involves IBM resources in addition to the original resources already being used to solve the issue. CritSits are perceived as poor management of PMRs, regardless of the underlying cause. Avoiding and reducing CritSits are top priorities for IBM.

\subsection{The Problem}

Currently, support analysts are tasked with handling PMRs by responding to customer emails: answering questions and offering advice on how to get passed their issue. Manually tracking risk of escalation, however, requires detailed attention beyond the PMR itself and towards the customer behind the PMR. The support analyst can track the business and emotional state of the customer, and ultimately make judgment calls on whether they think a PMR is likely to escalate. This becomes tedious as support analysts manage more and more customers, as each customer within this ecosystem might be related to multiple products and support teams. Dissatisfaction with any of the other products might result in escalations by the customer; furthermore, customers inevitably have trends, repeat issues, and long term historical relationships that might contribute to escalations. To manage the tracking and predictive modelling of all PMRs in the IBM ecosystem, an automated solution was required.

\section{Support Ticket Model Features (RQ1)}
\label{sec:stmf_features}

\begin{table*}[ht]
\centering
\caption{Support Ticket Model Features with Stages of Development}
\label{table:final_features_1}
\resizebox{\textwidth}{!}{%
\def\arraystretch{1.2}
\begin{tabular}{|c|l|l|c|c|c|}
    \hline
        \multicolumn{1}{|c|}{\multirow{3}{*}{\textbf{Category}}}
        & \multicolumn{1}{c|}{\multirow{3}{*}{\textbf{Feature}}}
        & \multicolumn{1}{c|}{\multirow{3}{*}{\textbf{Description}}}
        & \multicolumn{3}{c|}{\textbf{\nlcell{c}{Created or Improved During}}}
        \\\cline{4-6} & & &
        \textbf{\nlcell{c}{Problem \\ Characterization}}
        % & \textbf{\nlcell{c}{Eval 1}}
        % & \textbf{\nlcell{c}{Eval 2}}
        & \textbf{\nlcell{c}{Evalua- \\ tion 1}}
        & \textbf{\nlcell{c}{Evalua- \\ tion 2}}
    \\\hline\hline
        \multirow{3.5}{*}{\nlcell[1.2]{c}{Basic\\ Attributes}}
            & Number of entries
            & Number of events/actions on the PMR
            & \checkmark & &
        \\\cline{2-6}
            & Days open
            & Days from open to close (or CritSit)
            & \checkmark & &
        \\\cline{2-6}
            & PMR ownership level
            & \nlcell{l}{Level of Support (L0 - L3) that is in charge \\ of the PMR, calculated per entry}
            & & & \checkmark 
    \\\hline\hline
        \multirow{5}{*}{\nlcell[1.2]{c}{Customer\\ Perception\\ of Process}}
            & \nlcell{l}{Number of support people \\ in contact with customer}
            & \nlcell{l}{Number of support people the \\ customer is communicating with}
            & \checkmark & &
        \\\cline{2-6}
            & Number of increases in severity
            & Number of times the Severity increase
            & \checkmark & &
        \\\cline{2-6}
            & Number of decreases in severity
            & Number of times the Severity decrease
            & \checkmark & &
        \\\cline{2-6}
            & \nlcell{l}{Number of sev4/sev3/sev2 \\ to sev1 transitions}
            & \nlcell{l}{Number of changes in Severity\\ from 4, 3, or 2, straight to 1}
            & \checkmark & &
    \\\hline\hline
        \multirow{7}{*}{\nlcell[1.2]{c}{Customer\\ Perception\\ of Time}}
            & Time until first contact
            & \nlcell{l}{Minutes before the customer hears from\\ IBM for the first time on this PMR}
            & \checkmark & &
        \\\cline{2-6}
            & Current received response time
            & \nlcell{l}{Average number of minutes of all the\\ analyst response times on this PMR}
            & \checkmark & &
        \\\cline{2-6}
            & \nlcell{l}{Difference in current vs historical \\ received response time}
            & \nlcell{l}{(Historical received response time) minus \\ (Current received response time)}
            & \checkmark & &
        \\\cline{2-6}
            & Days since last contact
            & Number of days since last contact, calculated per entry
            & & & \checkmark
        \\\cline{2-6}
            & \nlcell{l}{Difference in historical sent vs \\ historical received response time}
            & \nlcell{l}{(Historical received response time) minus \\ (Historical sent response time)}
            & & & \checkmark
    \\\hline\hline
            & \multicolumn{1}{c|}{\nlcell[1.0]{c}{Decay of Information * \\ Live Indicators $\dagger$}} & & & &
    \\\hline\hline
        \multirow{8}{*}{\nlcell[1.0]{c}{Customer\\ Profile}}
            & Number of open PMRs * $\dagger$
            & Number of PMRs owned by customer that are open
            & & \checkmark &
        \\\cline{2-6}
            & Number of closed PMRs *
            & Number of PMRs owned by customer that are closed
            & \checkmark & \checkmark &
        \\\cline{2-6}
            & Number of open CritSits * $\dagger$
            & Number of CritSits owned by customer that are open
            & & \checkmark &
        \\\cline{2-6}
            & Number of closed CritSits *
            & Number of CritSits owned by customer that are closed
            & \checkmark & \checkmark &
        \\\cline{2-6}
            & Open CritSit to PMR ratio $\dagger$
            & (Number of open CritSits) / (Number of open PMRs)
            & & \checkmark &
        \\\cline{2-6}
            & Closed CritSit to PMR ratio
            & (Number of closed CritSits) / (Number of closed PMRs)
            & \checkmark & &
        \\\cline{2-6}
            & Historical received response time
            & \nlcell{l}{Average of all received response times \\ on PMRs owned by this customer}
            & \checkmark & &
    \\\hline\hline
        \multirow{7}{*}{\nlcell[1.0]{c}{Support\\ Analyst\\ Profile}}
            & Number of open PMRs * $\dagger$
            & Number of PMRs owned by customer that are closed
            & & & \checkmark
        \\\cline{2-6}
            & Number of closed PMRs *
            & Number of PMRs owned by the analyst that are closed
            & & & \checkmark
        \\\cline{2-6}
            & Number of open CritSits * $\dagger$
            & Number of CritSits owned by the analyst that are open
            & & & \checkmark
        \\\cline{2-6}
            & Number of closed CritSits *
            & Number of CritSits owned by the analyst that are closed
            & & & \checkmark
        \\\cline{2-6}
            & Open CritSit to PMR ratio $\dagger$
            & (Number of open CritSits) / (Number of open PMRs)
            & & & \checkmark
        \\\cline{2-6}
            & Closed CritSit to PMR ratio
            & (Number of closed CritSits) / (Number of closed PMRs)
            & & & \checkmark
        \\\cline{2-6}
            & Historical sent response time
            & \nlcell{l}{Average of all sent response times \\ on PMRs owned by an analyst}
            & & & \checkmark
    \\\hline
        \multicolumn{6}{r}{\nlcell{c}{* in the last N weeks, where N = $\infty$, 12, 24, 36, and 48}}
\end{tabular}%
}
\end{table*}

Table \ref{table:final_features_1} outlines the Support Ticket Model Features created during this research. This table reflects the final set of features that was used in the final model, producing the final set of results. For each feature, we provide a brief description, as well as a marker identifying at which stage of the design science methodology each feature was created or improved. The ``Created or Improved During" column has three sub-options: ``Problem Characterization" features were created immediately following the Problem Characterization phase, ``Eval 1" features were created or improved following the Evaluation 1 phase, and ``Eval 2" features were created or improved following the Evaluation 2 phase.

\subsection{Basic Features}
The features in this category are characterized by their immediate availability in offering value to the Support Ticket Model Features without any modification from the state in which IBM maintains them. When support analysts are addressing PMRs, the \textit{Number of entries} represents how many actions or events have occurred on the PMR to date (e.g. an email is received, a phone call is recorded, the severity increased, etc.). Lastly, the number of \textit{Days open} keeps track of days since the PMR was opened. Finally, \textit{PMR ownership level} tracks the different levels of support that a PMR can be at, starting from L0 up to L3 (detailed in Section \ref{ssec:ethno}).

\subsection{Customer Perception of Process}
The features in this category are characterized by the perspective they offer in harnessing the customer's perception of the support process as a separate experience from the way in which support analysts perceive the support process. The customer's perspective of process can be engineered using data that is visible to them and ignoring data that is not. If a customer wants to convey the urgency or importance of their issue, the severity field on their PMR is the way to do that; customers are in charge of setting the severity of their PMRs. Severity is a value from 4 to 1, with 1 being the most severe; severity can be changed to any number at any time. Any \textit{Number of increases in severity} is a sign that the customer believes their issue is becoming more urgent; conversely, any \textit{Number of decreases in severity} can be interpreted as the issue improving. Support analysts watch for increases to severity, but the most severe situations are modelled by the \textit{Number of sev4/sev3/sev2 to sev1 transitions}, as this represents the customer bringing maximum attention to their PMR. Finally, within the support process, there are many people involved with solving customer issues, but there are only a certain \textit{Number of support people in contact with the customer}.

\subsection{Customer Perception of Time}
Similarly, the customer's perception of time can be engineered using timestamps and ignoring PMR activity that is not visible to the them. The first time when customers may become uneasy is the \textit{Time until first contact} with a support analyst. At this stage the customer is helpless to do anything except wait, which is a unique time in the support process. Once a customer is in contact with support there is an ongoing back-and-forth conversation that takes place through emails and phone calls, the timestamps of which are used to build the \textit{Current received response time}. Each customer has their own expectation of response time from their historical experience with IBM support, which in turn can be compared to the current received response time. This \textit{Difference in current vs historical received response time} requires that the customer's historical received response time is known, which is explained in the next feature category. \textit{Days since last contact} was introduced as a feature because this is one of the most important factors to IBM in maintaining constant communication with their customers. This feature represents how many days it has been since contact has been made between the customer and support. Finally, \textit{Difference in historical sent vs historical received response time} is a feature that highlights the difference between what the customer expects from support given their historical experiences of receiving responses from support, against what the analyst is likely to send as a response time given their historical sent response times.

\subsection{Customer Profile}
The features in this category harness historical information about customers as entities within the support organization, spanning across all support tickets they have ever opened. Tracking customer history allows for insights into customer-specific behaviors that manifest as trends across their PMRs. The customer is the gate-keeper of information, the one who sets the pace for the issue, and the sole stakeholder who has anything to gain from escalating their PMR. As such, it seems appropriate to model the customer over the course of all their support tickets. Customers within the IBM ecosystem have a \textit{Number of closed PMRs} and a \textit{Number of closed CritSits}. Combined, these two numbers create a \textit{Closed CritSit to PMR ratio} that represents the historical likelihood that a customer will Crit their future PMRs. Customers also have a \textit{Historical received response time} from their past experiences with IBM support. This is calculated by averaging the ``Current received response time" feature over all PMRs owned by a customer. Finally, the customer has a \textit{Number of open PMRs} and a \textit{Number of open CritSits} that together reflect the current state of the customers support experience, captured in the combined feature \textit{Open CritSit to PMR ratio}. As marked in Table \ref{table:final_features_1}, the features in this category have two sub-groups that define them: Decay of Information and Live Indicators. ``Decay of Information" features only retain information for a set period of time, as reflected in the names of the features. An example of this is ``Number of closed PMRs," which later becomes five separate features, one of which is ``Number of closed PMRs in the last 12 weeks." This feature reflects how many PMRs this customer has closed in the last 12 weeks, which is different than the other four features which all have a different number of weeks. ``Live Indicators" features harness support tickets and escalation artifacts that were open when the target PMR was open. For example, ``Number of open PMRs" reflects how many PMRs (owned by the same customer) were open when the target PMR Crit or closed, thereby creating an indicator of a live (real-time) part of the data.

\subsection{Support Analyst Profile}
Similar to the Customer Profile category, features in this category harness historical information about support analysts as entities within the support organization, spanning across all support tickets they have handled. During the lifetime of a PMR, a number of support analysts may contribute to the overall solution delivered to the customer; however, there will be one support analyst who contacts the customer more than any other support analyst, and they are tagged as the lead support analyst for that PMR. Within IBM's support ecosystem, that support analyst has accumulated a \textit{Number of closed PMRs} and a \textit{Number of closed CritSits} over time. At any one time, they also have a \textit{Number of open PMRs} and a \textit{Number of open CritSits}. Both the open and closed states of the support analyst's experience are summed up in the features \textit{Closed CritSit to PMR ratio} and \textit{Open CritSit to PMR ratio}. Finally,  across all of those PMRs, the \textit{Historical sent response time} of an analyst can be calculated by averaging all of their response times to customers across all PMRs. Similar to the Customer Profile category, features in the Support Analyst Profile have two sub-groups: Decay of Information and Live Indicators.

\section{Engineering the Features in the Support Ticket Model (RQ2)}

Our approach to addressing the manual process of tracking PMRs and their escalations began by modeling PMR information available to analysts in assessing the possibility of a customer escalating their issue, followed by engineering the Support Ticket Model Features (RQ1). To begin the FE process, we analyzed data from our on-site observations and conducted further interviews aimed specifically at understanding how analysts reason through the information about their PMRs and customers. We first describe the interview questions and data we gathered, followed by our data analysis procedure.

\subsection{Interviews}

We conducted a series of semi-structured interviews with support analysts at IBM, five at IBM Victoria and four in worldwide customer support organizations, all of whom are customer-facing in their daily jobs. We sought to identify information that is currently available in customer records and support tickets, particularly information analysts use to assess the risk of support ticket escalations. We asked questions such as ``Why do customers escalate their issues?", ``Can you identify certain attributes about the issue, customer, or IBM that may trigger customers to escalate their issue?", as well as exploratory questions about support ticket data as we identified in the PMR repository. The full interview script can be found online \footnote{\url{http://thesegalgroup.org/wp-content/uploads/2017/02/support-analyst.pdf}}.

\subsection{Thematic Analysis}
Thematic analysis \cite{Cruzes2011} was used to analyze the interview transcripts. We labelled the responses with thematic codes that represented possible directions for ML features that could automate the process of CritSit prediction. From there we grouped the codes into thematic themes, which later became the feature categories. The themes and underlying codes are listed in Table \ref{table:thematic}. We validated and refined these themes and codes through two focus groups consisting of: the Victoria Site Manager, the L3 Support Analyst, and an L2 Support Analyst.

\begin{table}[hb]
\centering
\caption{PMR-Related Information from Interviews, Relevant to Predicting PMR Escalations }
\label{table:thematic}
\resizebox{\columnwidth}{!}{\def\arraystretch{1.2}
\begin{tabular}{|l|l|}
    \hline
        \textbf{Themes}
        & \textbf{Codes}
    \\\hline
        Basic Features
        & How long has a PMR been open
    \\\hline
        \multirow{2}{*}{\nlcell{l}{Customer Perception \\ of the PMR Process}}
            & Fluctuations in severity
            \\\cline{2-2} 
            & Support analyst involvement
    \\\hline
        \multirow{2}{*}{\nlcell{l}{Customer Perception \\ of Time with Respect \\ to their PMR}}
            & Initial response wait time
            \\\cline{2-2} 
            & \nlcell{l}{Average response wait time \\ on respective PMRs}
    \\\hline
        \multirow{3}{*}{Traits of Customers}
            & How many PMRs they have owned
            \\\cline{2-2} 
            & How many CritSits they have owned
            \\\cline{2-2} 
            & Expectation of response time
    \\\hline
\end{tabular}}
\end{table}

\subsection{A First Set of Features in the Support Ticket Model}
\label{ssec:original_model}

To develop the Support Ticket Model Features, we mapped PMR repository data to the codes from our analysis under each of the themes we identified, creating the first 13 Support Ticket Model Features (see Table \ref{table:final_features_1}, marked under ``Problem Characterization"). The number of features and the features themselves emerged during the thematic analysis analysis of our Problem Characterization stage.

Throughout this process certain types of PMR data were usable as-is, without modifying the data in IBM's dataset such as ``number of days open", and other types of data had to be restructured, counted, or averaged. An example of a more complicated mapping is the ``Number of open PMRs" which, conceptually, is a feature that at any time should reflect how many PMRs a customer has open. However, to actually create this feature for a PMR involves identifying the customer and picking a point in time, followed by implementing an algorithm to go through all PMRs to figure out which ones are owned by that customer and between the open and close dates that match the chosen point in time. The ``point in time" chosen for PMRs is the moment before the CritSit occurs, or the moment before it closes (if the PMR does not Crit).

Once a code had data mapped to it, it was considered a feature of the model. In developing the model features, we sought to abstract as much as possible from the specifics of IBM's data and processes to increase transferability to other organizations. Our approach to achieve transferability to other organizations was to generalize or remove features that were not broad enough to the support process in general that other organizations were likely to be able to implement them. This approach requires knowledge of support processes in ``other organizations," of which two of the involved researchers had, as well as a small number of the interviewed senior managers at IBM who had spent time at other organizations.
\section{Evaluation 1: In-Depth Review of the Support Ticket Model with IBM Analysts}
\label{sec:eval1}

Our first evaluation sought to validate the first set of features in our Support Ticket Model with IBM. In order to do that, however, the features had to be used in a ML algorithm to produce results that could be reviewed (RQ2). We evaluated the output of the ML model through statistical validation as well as with IBM support analysts at multiple sites.

\subsection{Machine Learning Model}

The creation of the ML model was straightforward once PMR data had been mapped to the first set of features in the Support Ticket Model. We fed the 13 Support Ticket Model Features into multiple supervised ML algorithms: CHAID \cite{McCarty2007}, SVM \cite{Pang-Ning2006}, Logistic Regression \cite{hosmer2013applied}, and Random Forest \cite{Pang-Ning2006}. Although other algorithms produced higher precision, we chose Random Forest because it produced the highest recall. High recall was preferred for two reasons: as argued by Berry \cite{Berry2017} and exemplified in the recent work of Merten et al. \cite{Merten2016}. Additionally, our industrial partner expressed a business goal of identifying problematic PMRs while missing as few as possible. The input we received from the IBM analysts was that they would prefer to give more attention to PMRs that have potential to Crit, rather than potentially missing CritSits. In other words, they were more comfortable with false positives than false negatives.

The Random Forest model we built has a binary output, as the input of our target class is 0 or 1. Random Forest outputs a confidence in each prediction, which we correlated to the PMR's risk of escalation, or Escalation Risk (ER). For example, if the model output a prediction of 1, with confidence 0.88, this PMR's ER is 88\%. Any ER over 50\% is categorized as a Crit.

The ratio of CritSit to non-CritSit PMRs is extremely unbalanced at 1:250, therefore some kind of balancing was required to perform the ML task. The Random Forest classifier we used has the capability to handle imbalanced data using oversampling of the minority class \cite{Tan2006}. In other words, the algorithm re-samples the minority class (CritSit) roughly enough times to make the ratio 1:1, which ultimately means that each of the minority class items are used 250 times during the training phase of the model. This method allows all of the majority class items to be used in learning about the majority class, at the cost of over-using the minority items during the learning phase.

\subsection{Statistical Results \& Validation - First Features}
\label{ssec:statistical_results_original}

All PMRs and CritSits were randomly distributed into 10 folds, and then 10-fold leave-one-out cross-validation was performed on the dataset using the Random Forest classifier. The results of the validation can be seen in the confusion matrix in Table \ref{table:conf_matrix_original_rf}. A confusion matrix is a useful method of analyzing classification results \cite{Fawcett2004} that graphs the True Positives (TP), True Negatives (TN), False Positives (FP), and False Negatives (FN). The diagonal cells from top-left to bottom-right represent correct predictions (TN and TP).

The recall for ``CritSit -- Yes" is 79.94\%, with a precision of 1.65\%. Recall and precision are calculated as $\frac{TP}{TP + FN}$ and $\frac{TP}{TP + FP}$, respectively. The recall of 79.94\% means that the model is retrieving 79.94\% of the relevant PMRs (CritSits), whereas the precision of 1.65\% means that the algorithm is retrieving a lot more Non-CritSit PMRs than CritSit PMRs, so much so that the ratio of CritSit PMRs to all PMRs retrieved is 1.65\%. 

As previously mentioned, our business goal for building the predictive model was to maximize the recall. Additionally, Berry et al. \cite{Berry2012} argue about tuning models to predict in favor of recall when it is generally easier to correct FPs than it is to correct TNs. Significant work has been completed towards identifying which of the PMRs are CritSits, this work is measured through the metric ``summarization", calculated as $\frac{TN + FN}{TN + FN + TP + FP}$. In short, summarization is the percentage of work done by classification algorithms towards reducing the size of the original set, given that the new set is the sum of FP + TP \cite{Berry2017}. Summarization alone, however, is not useful, it must be balanced against recall. 100\% recall and any summarization value greater than 0\% is progress towards solving identification and classification problems. Our model has 79.94\% recall and 80.77\% summarization. Simply put, if a support analyst wanted to spend time identifying potential CritSits from PMRs, our model reduces the number of candidate PMRs by 80.77\%, with the statistical guarantee that 79.94\% of CritSits remain.

\begin{table}[b]
\centering
\caption{Confusion Matrix for CritSit Prediction using Random Forest on First Features}
\label{table:conf_matrix_original_rf}
\resizebox{\columnwidth}{!}{
    \begin{tabular}{|c|c|c|c|}
        \hline
            \multirow{2}{*}{Actual}
            & \multirow{2}{*}{Total}
            & \multicolumn{2}{c|}{Predicted as}
            \\\cline{3-4} 
                & & CritSit - No
                & CritSit - Yes
        \\\hline
            CritSit - No
            & 2,557,730
            & \nlcell{c}{2,072,496 (TN) \\ 81.03\%}
            & \nlcell{c}{485,234 (FP) \\ 18.97\%}
        \\\hline
            CritSit - Yes
            & 10,199
            & \nlcell{c}{2,046 (FN) \\ 20.06\%}
            & \nlcell{c}{8,153 (TP) \\ 79.94\%}
       \\\hline
    \end{tabular}
}
\end{table}

\subsection{Model Output Evaluation}
\label{ssec:model_output_validation}

Using our close relationship with IBM Victoria, we then conducted an in-depth review of the model output in a 2-hour meeting with the support analysts and managers, to gain deeper insights into the behavior of the model on an individual PMR-level basis, to improve the model features.

\subsubsection{Evaluation Setting}

We examined ten major (suggested by IBM) closed CritSit PMRs from IBM Victoria in our dataset and ran our ML model to produce escalation-risk graphs for each of the CritSit PMRs. The ten CritSit PMRs chosen by IBM were memorable escalations, memorable enough to be discussed with clarity. We show six of the ten graphs in Figs. \ref{fig:pmr_er_1}-\ref{fig:pmr_er_3}, each graph is a single PMR. The graphs plot the ER as produced by our ML model over time, from the first snapshot to its last snapshot. By ``snapshot" we are referring to the historical entries that exist per PMR. E.g., a PMR with 16 changes to its data will have 16 snapshots, each consecutive snapshot containing the data from the last snapshot plus one more change. Our goal was to compare the output of our model with what IBM remembered about these ten PMRs when they were handled as escalating issues (i.e. at the time of each snapshot).

The 2-hour in-depth review involved four IBM support representatives: the Site Manager, the Development Manager, the L3 Support Analyst, and an L2 Support Analyst. We printed the graphs of these ten CritSit PMRs, discussed them as described below, and took notes during the meeting:

\begin{enumerate}[a.]
    \item Revealing to the members PMR numbers and customer names of the PMRs in the analysis, allowing them to look up these PMRs in their system and read through them.
    \item Discussed the PMRs in the order the members preferred.
    \item Displayed the graphs of the Escalation Risks.
    \item Inquired about how the model performed during each PMR in comparison to what they experienced at the time.
\end{enumerate}

\subsubsection{Evaluation Results}
\label{sssec:e1_findings}

Overall, our ML model performed well in predicting the ER per PMR, per snapshot. However, the findings of this in-depth review of the model are broader and pertain to a) improvements in our model with respect to the Customer Profile information and b) our increased understanding of IBM's support process. Both findings relate to refinements in our model as well as recommendations to other organizations intending to apply our model to perform EP.

\begin{figure}[t]
    \centering
    
    \includegraphics[width=\columnwidth]{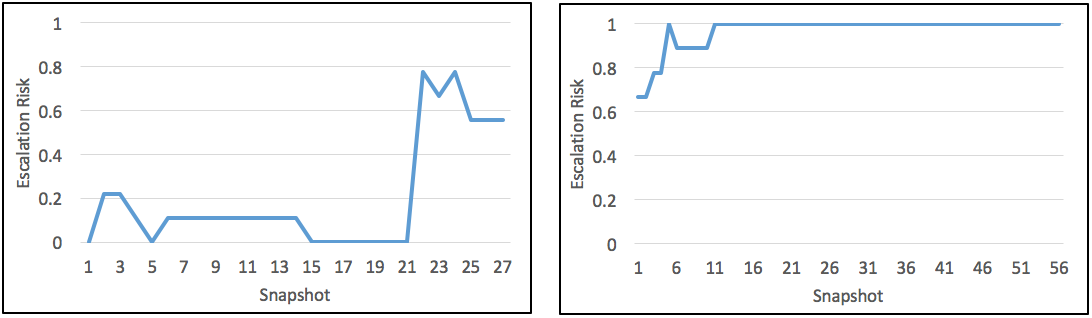}
    \caption{Two PMRs with little-to-no Customer Profile info built ER over time}
    \label{fig:pmr_er_1}
    \vspace{8mm}

    \includegraphics[width=\columnwidth]{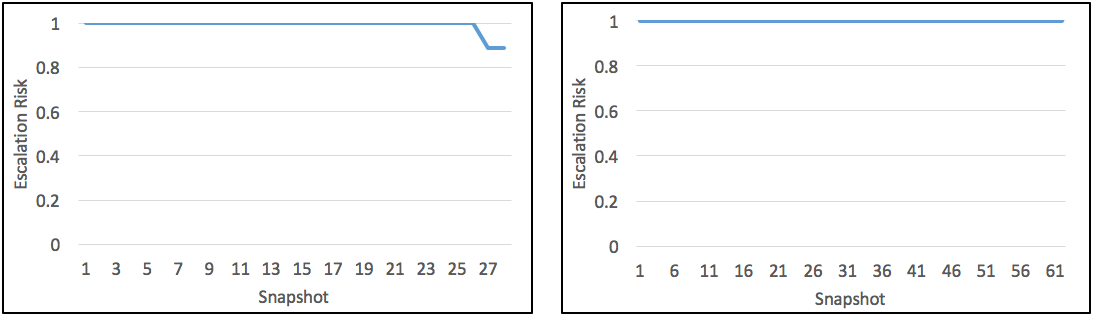}
    \caption{Two PMRs with too much Customer Profile info defaulted to high ER early}
    \label{fig:pmr_er_2}
    \vspace{8mm}

    \includegraphics[width=\columnwidth]{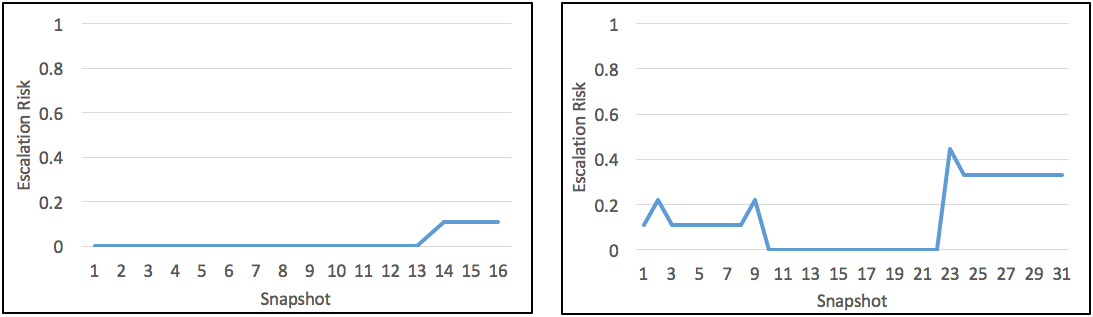}
    \caption{Two ``Cascade" CritSits showed low ER}
    \label{fig:pmr_er_3}
\end{figure}

\textbf{Role of Historical Customer Profile Information.} Two of the ten PMRs in this evaluation showed a trend of building ER over time as events occurred, as shown in Fig. \ref{fig:pmr_er_1}. Manual inspection and discussion with the analysts indicate that this behavior was correlated with a lack of Customer Profile information for both PMRs. All Customer Profile features (see Table \ref{table:final_features_1}) refer to data that is available when the PMR is created and will not change during the lifetime of the PMR; therefore, the initial ER is solely due to the Customer Profile features, and the changes in ER during the lifetime of the PMR must be due to the other categories.

In contrast, PMRs with too much Customer Profile information were immediately flagged as CritSits. The model had learned that excessive Customer Profile information correlates with high ER. Five of the ten PMRs had this behavior, two of which can be seen in Fig. \ref{fig:pmr_er_2}. Manual inspection of the five PMRs revealed a lot of Customer Profile information for each of the five PMRs, i.e., the ``Number of Closed PMRs" field was 200+ for each of the five customers of these PMRs.

These findings show variance in model performance for the two extremes of quantity of Customer Profile information in the PMRs we studied. We saw expected behavior for lack of Customer Profile information but unexpected behavior for the opposite, PMRs with extensive Customer Profile information. These variances point to the role of the Customer Profile category in capturing aspects of the customer beyond the current PMR, allowing traits of the customer to be considered during the prediction of escalation risk. To properly capture the features of the Customer Profile category, we made refinements to our model by adding new features that add decay of customer information over time, such that the history does not exist forever. These features are discussed in Section \ref{sssec:feeding_model_1}.

\textbf{Recording True Reason for CritSit PMRs is Important.} The second insight from this study was about IBM's support process and feedback into revised features in our model. We ran into a situation where on some of the PMRs our model showed low ERs, although they appeared officially as CritSits in the IBM system. Through manual inspection of PMR historical information, our study participants identified that these PMRs were not the cause of the CritSit, and in fact there were other PMRs with the same CritSit ID that were responsible for them being recorded as CritSits in the IBM system. We discovered that it is common practice to Crit every PMR owned by a customer when any one of their PMRs Crit. Therefore, there was a distinction between the ``cause" CritSit --- the CritSit PMR that caused the Crit to happen, and ``cascade" CritSits --- the CritSit PMR(s) that subsequently Crit due to the process of applying a Crit to every PMR owned by the same Customer in response to some ``cause" CritSit. Figure \ref{fig:pmr_er_3} shows two of the three PMRs that had this behavior (``cascade" CritSits) in which our model behaved correctly.

Evaluation 1 led to the Customer Profile feature category receiving new and modified features. The new features, denoted as ``Decay of Information", forget information over time to give current information more influence. The modified features, listed in Table \ref{table:final_features_1} and marked under the column ``Eval 1" are \textit{Number of PMRs Closed in the last N weeks} and \textit{Number of CritSits closed in the last N weeks}, where ``N" is infinity, 12, 24, 36, and 48. Prior to this phase of the research, they did not have the ``in the last N weeks" ending. In addition to the above changes, the suggestion to track the decay of information lead to the observation that we were not tracking ``now" in the sense of what else is open while a PMR is active. In other words, if a customer has a PMR that escalates into a CritSit, did this customer have other open PMRs that may affect their decision to escalate? Did they have other open CritSits? These inquires lead to the new features \textit{Number of PMRs opened in the last N weeks} and \textit{Number of CritSits opened in the last N weeks}, which incorporates the new feature category ``Live Indicators" as well as the previous ``Decay of Information". This new perspective on the archival data provides the ML algorithm with the option to utilize smaller and more recent subsets of the entire history so that recent events are not overshadowed by past events.

\subsubsection{Feeding Back into the Model}
\label{sssec:feeding_model_1}

Evaluation 1 lead to the creation of new features (and modifying existing features) under the existing feature category, Customer Profile, and two new feature categories, Decay of Information and Live Indicators.

Decay of Information features have a quantiﬁer attached that dictates how many weeks they retain information: infinite, 12, 24, 36, and 48. These features are marked with a ``*" in Table 1. The other new feature category created is ``Live Indicators", denoted with a ``$\dagger$" in Table 1. These features capture the number of PMRs and CritSits that a customer had open when dealing with their PMR.

\section{Evaluation 2: In Situ Evaluation with Support Analysts}
\label{sec:eval2}

The second evaluation investigated the assistance provided by our model running in real time during the management meetings at the Victoria site when analysts together with management discussed open PMRs. To do this, we developed a prototype tool \cite{Montgomery2017Tool} that displays all open PMRs and their current predicted ER, as well as the 13 first features –-- per PMR –-- that go into the prediction.

\subsection{Our Prototype}

\begin{figure*}[ht]
    \centering
    
    \includegraphics[width=\textwidth]{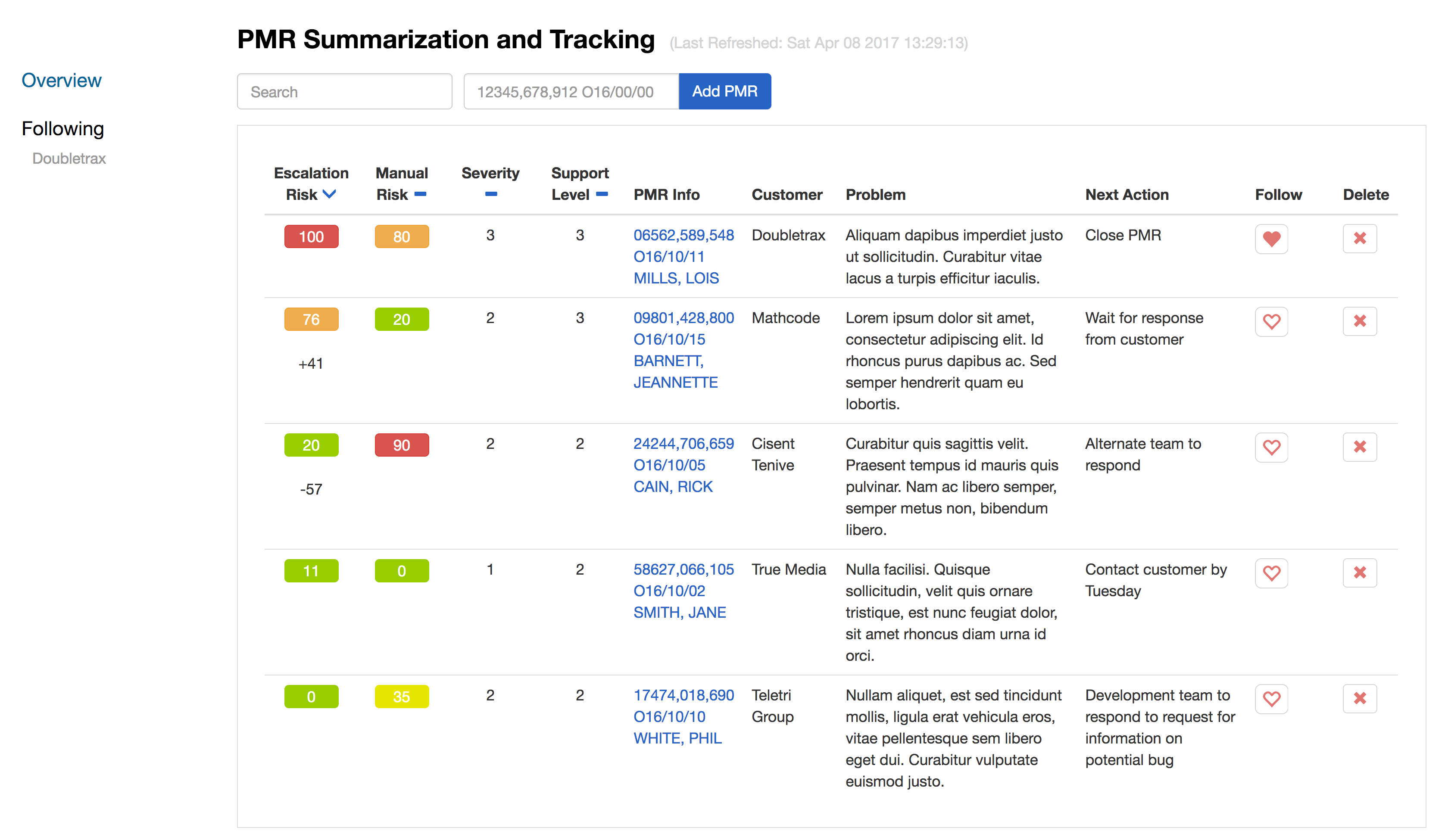}
    \caption{Prototype Tool Overview Page}
    \label{fig:tool_overview}
\end{figure*}

\begin{figure*}[ht]
    \centering
    \includegraphics[width=\textwidth]{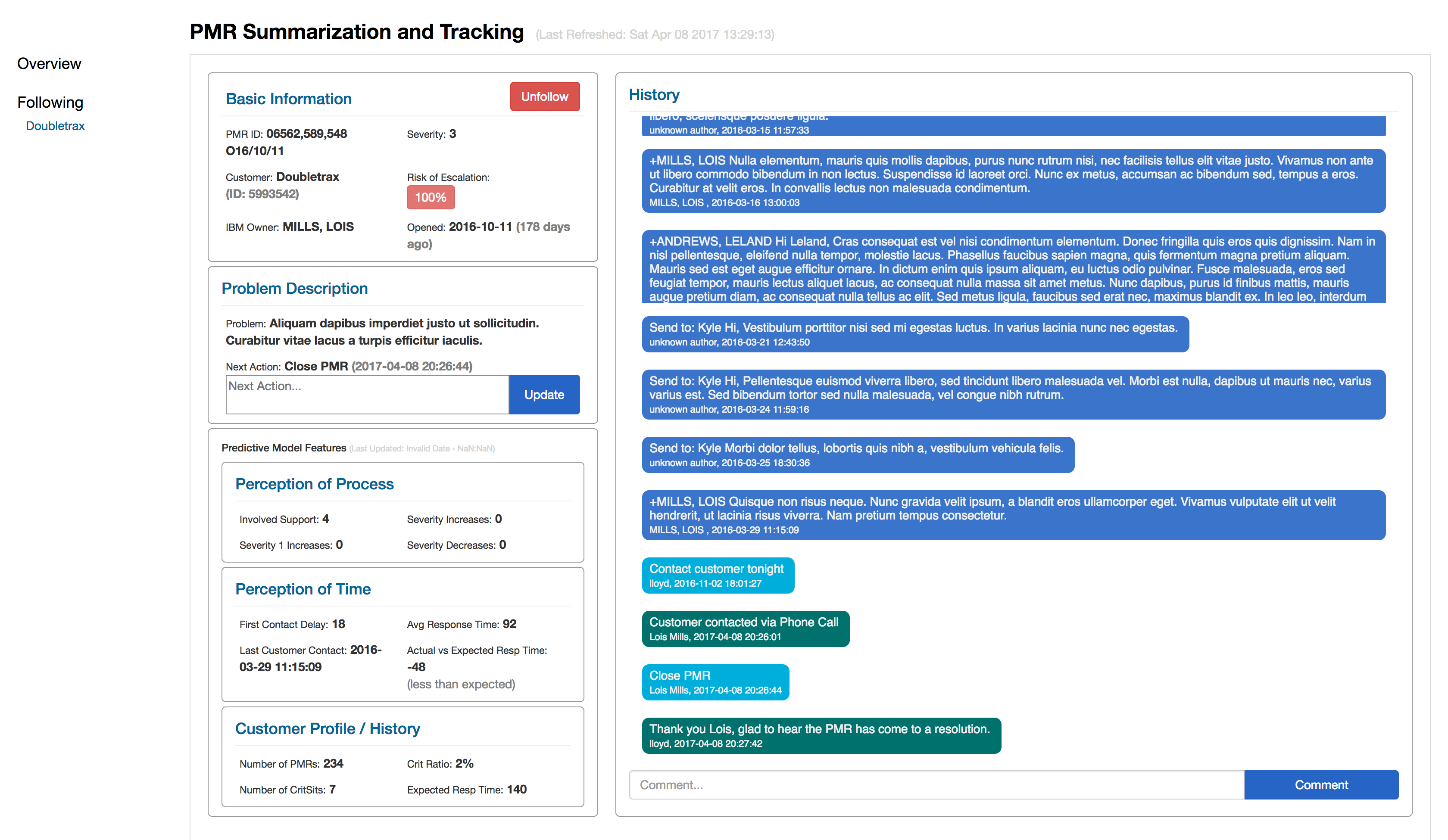}
    \caption{Prototype Tool In-depth Page}
    \label{fig:tool_indepth}
\end{figure*}

Our prototype tool displayed all active PMRs at the Victoria site with two main displays: the overview, and the in-depth view. The overview displays all open PMRs in a summarized fashion for quick review (Fig. \ref{fig:tool_overview}). The in-depth view comes up when a PMR is selected and shows the details of the PMR (Fig. \ref{fig:tool_indepth}). Included in this view is: the history of email correspondence between support and customer, description of the issue, and the ML model features that were used to produce the ER.

\subsection{Evaluation Setting}

We evaluated the use of our prototype over a period of four weeks during daily stand-up support meetings with managers and support analysts. Prior to this tool, these stand-up meetings were managed day-to-day by an excel sheet stored locally on the Site Manager's computer. The effectiveness of the meetings relied on support analysts to bring up and discuss PMRs they were working on.

Our prototype was first evaluated in a pilot study, to gain feedback on shortfalls and bugs. After the short (one week) pilot, a week was spent improving the tool based on recommendations before the full four-week deployment. The participants of this study were the Victoria Site Manager, the Development Manager, the L3 Support Analyst, and two L2 Support Analysts. One of the researchers participated in all these meetings while the tool was in use for the first two weeks of the study, as well as two days near the end of the study.

After the pilot study, two additional features were added to the tool: (1) Displaying a Manual Escalation Risk (MER), a number field from 0 to 100 (to be input by anyone on the team) to eliminate the need to remember the analysts' assessments of each PMR during past meetings; and (2) Displaying a Change in Escalation Risk (CER), a number field from -100 to 100 that represents the change in ER since the last update, to eliminate the need for anyone to memorize ERs by tracking changes manually. With the MER and CER being tracked and displayed, the team could expedite the daily PMR review process and focus on PMRs that either had a high MER or CER.

\subsection{Evaluation Findings}

The use of our prototype during the PMR management meetings allowed them to focus on the PMRs that had greater potential to escalate. In the absence of our tool, the analysts would review PMRs brought up by support analysts and discuss them based on the memory of the participants, often relying on management to bring up additional items they had forgotten. With our tool, they were able to parse through a list of PMRs ranked by ER. The MER capability allowed them to record their own assessment of the ER, and compare it with the ER output by our ML model. It allowed for subsequent meetings to be quicker because the team could see their past evaluations of PMRs, and focus on ones they had assigned a high MER. The CER field provided a quick reference to which PMRs had increased in ER since the last update.

During the evaluation period, the team identified that there were two important aspects of PMRs that mattered to them as well as the customer: PMR ownership level, and days since last contact. PMRs are always being directly managed by some level of support, and the difference between L2 and L3 support means a lot to IBM as well as the customer. L2 is product-usage support, where customers are generally at fault, and L3 is development-level support, where bugs are triaged and the product is at fault. Similarly, the number of days since last customer contact was brought up as an important factor for deciding when a customer may Crit. As a result of these discussions, two new features were added to our final set of model features in Table \ref{table:final_features_1}: \textit{PMR ownership level} and \textit{Days since last contact}.

Another finding that arose during this evaluation was that our model had no information regarding support analysts. A PMR largely involves two stakeholders: the customer, and the support analyst. Therefore, capturing some archived characteristics of the support analyst working on the PMR became a new category of features called ``Support Analyst Profile" as shown in Table \ref{table:final_features_1}. The features in this category closely mirror those of the Customer Profile category, except from the perspective of a particular support analyst, instead of a particular customer.

\subsubsection{Feeding Back into the Model}

This evaluation cycle produced new features in our Support Ticket Model under the existing feature categories Basic Attributes and Customer Perception of Time, as well as under new feature category ``Support Analyst Profile". 

The Basic Attributes feature category received the new feature ``PMR Ownership Level" which reflects which level of support is currently handling the PMR (L0, L1, L2, or L3). Customer Perception of Time received ``Days since last contact", which reflects how long it has been since support contacted the customer, and ``Difference in historical sent vs historical received response time", which reflects the difference between what the customer has historically received as a response time and what the analyst has historically sent as a response time. The new feature category Support Analyst Profile was created to mimic the features under the Customer Profile category, except from the perspective of the support analyst. The Support Analyst Profile has four features that incorporate Decay of Information qualifiers, and three features that fall under Live Indicators.

\section{Evaluation 3: Additional Feature Engineering \& Statistical Validation of Final Model}
\label{sec:eval3}

For this evaluation, changes were made to algorithms being used, additional FE was conducted, and all model features, including those developed through the two rounds of evaluations, were validated using statistical methods. 

\subsection{Switching from Random Forest to XGBoost}
\label{ssec:xgboost}

Based on a suggestion during the previous evaluation cycles, XGBoost was tried in place of Random Forest as the ML algorithm for this research. The results for each algorithm are comparable at the previously mentioned confidence threshold of 50\%, however, further investigation showed promising evidence towards switching to XGBoost.

XGBoost is a ML algorithm that, similar to Random Forest, uses tree structures to store the internal state of the model  \cite{Chen2016}. However, XGBoost produced a more diverse Precision-Recall Space (PR Space) than Random Forest. The standard way to compare ML implementations is the Receiver Operating Characteristic (ROC) graph which plots the true positive rate against the false positive rate. However, we found in working closely with IBM that PR Space graphs were easier to explain and still allowed for decisions to be made about the models and their confidence thresholds. PR Space shows the trade-off in precision and recall that happens as confidence thresholds are changed and is noted ``as an alternative to ROC curves for tasks with a large skew in the class distribution" \cite{Davis2006}.

\begin{figure}[t]
    \centering
    \includegraphics[width=\columnwidth]{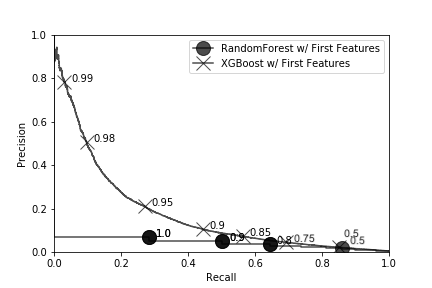}
    \caption{Random Forest vs XGBoost in PR Space (marked with confidence thresholds)}
    \label{fig:prs_rf_vs_xgb}
\end{figure}

Figure \ref{fig:prs_rf_vs_xgb} is a PR Space graph showing the difference between Random Forest and XGBoost in precision and recall across all confidence thresholds. The axes are labelled with ``Precision" and ``Recall", and the lines are labelled at various points with the confidence threshold at that point. To show how comparable the XGBoost results are to the Random Forest confusion matrix in Table \ref{table:conf_matrix_original_rf}, the results of using XGBoost with the first features are detailed in Table \ref{table:conf_matrix_original_xgb}. The same first features are being used in both implementations, but the number of PMRs is reduced because during the evaluation cycles we identified PMRs with an issue that disqualified them from the analysis. The reduced dataset lost less than 1\% of the original data, and has an imbalance of 1:265.

\begin{table}[b]
\centering
\caption{Confusion Matrix for CritSit Prediction using XGBoost on First Features, with confidence threshold of 50\%}
\label{table:conf_matrix_original_xgb}
\resizebox{\columnwidth}{!}{
    \def\arraystretch{1.2}
    \begin{tabular}{|c|c|c|c|}
        \hline
            \multirow{2}{*}{Actual}
            & \multirow{2}{*}{Total}
            & \multicolumn{2}{c|}{Predicted as}
            \\\cline{3-4} 
                & & CritSit - No
                & CritSit - Yes
        \\\hline
            CritSit - No
            & 2,532,745
            & \nlcell{c}{2,164,262 (TN) \\ 85.45\%}
            & \nlcell{c}{368,483 (FP) \\ 14.55\%}
        \\\hline
            CritSit - Yes & 9,536 &
            \nlcell{c}{1,417 (FN) \\ 14.86\%} &
            \nlcell{c}{8,119 (TP) \\ 85.14\%}
      \\\hline
    \end{tabular}
}
\end{table}

With Random Forest, there was little precision to be gained by changing the confidence threshold, and the recall had a drastic reduction at higher confidences. With XGBoost, there was the potential to get near 100\% precision if the confidence was tuned high enough, but still at the cost of a drastic reduction in recall. Although recall is still a high priority for this project and so tuning for high precision was not the objective, an algorithm that produces similar results for recall and also gives the option for much higher precision at higher confidences is preferred.

The ratio of CritSit to non-CritSit PMRs is unbalanced at 1:265, therefore some kind of balancing was required to perform the ML task. The XGBoost classifier can handle imbalanced data through cost sensitive learning, a technique that ``assigns the training examples of different classes with different weights, where the weights are in proportion to their corresponding misclassification costs" \cite{Liu2006}. The core concept is to mathematically force the model to care about CritSits by increasing the loss to the internal cost function if it fails to correctly predict them. In other words, false negatives were assigned a high penalty to discourage XGBoost from producing them, therefore encouraging more ``CritSit -- Yes" predictions which raises the TP rate as well as the FP rate. As previously mentioned, FPs were preferred over FN by our industry collaborator.

The overall impact to precision and recall is displayed in Fig. \ref{fig:prs_rf_vs_xgb}, but to provide comparable results to the Random Forest implementation, Table \ref{table:conf_matrix_original_xgb} is the confusion matrix of the results when the confidence threshold is set to 50\%.

\subsection{Engineering the Additional Features}
\label{ssec:additional_features}

This section details the engineering of the new features under two conceptual groups, Decay of Information and Live Indicators, and one new feature category, Support Analyst Profile.

\subsubsection{Decay of Information}

One of the findings from Evaluation 1 is that Customer Profile information plays a strong role in assessing whether or not a PMR will Crit, and that ``PMRs  with  too  much  Customer  Profile  information  were  immediately  flagged  as  CritSits" (Section \ref{sssec:e1_findings}). To address this issue, we integrated variables into the model that represented a decay of information over time so the ML model could better utilize this new perspective of the data.

Features such as ``Customer number of closed PMRs" would accumulate data indefinitely as the FE algorithms traversed the data. Incorporating variables that decay over time means to delete data as it becomes too far in the past. 

Instead of deleting that information completely, however, a number of variables are used to keep track of different time windows, so that the ML algorithm can decide what time window best correlates with the target class. For example, the feature ``Customer number of closed PMRs" reflects all PMRs ever closed by a particular customer, which may not be useful in understanding the recent history of the customer. To mitigate this ever increasing history, new features including ``Number of closed PMRs in the last 12 weeks" were created to provide different perspectives into the customer's history. The ``in the last 12 weeks" suffix reflects that this feature only contains historical information from the last 12 weeks. The full list of decayed features includes each of the features in Table \ref{table:final_features_1} with a ``*", with 12, 24, 36, and 48 weeks each, adding up to a total of 32 new features (infinity was already a feature).

\subsubsection{Live Indicators}

Past history plays a role in how customers and support analysts approach new PMRs and has been shown in Section \ref{sssec:e1_findings} to play an important role in predicting CritSits, so the next step in engineering features was to leverage the live artifacts that exist in IBM's ecosystem to create a number of \textit{live indicators}.

Two new features were engineered, detailed in Table \ref{table:final_features_1} marked under column ``Eval 1". These new features are \textit{Customer number of open PMRs}, \textit{Customer number of open CritSit PMRs}, and \textit{Open CritSit to PMR ratio}.

\subsubsection{Support Analyst Profile}

The second evaluation phase revealed the underlying importance of the Customer Profile features, which lead to the decision to incorporate another profile-like category: Support Analyst Profile. As such, the features in Table \ref{table:final_features_1} marked in column ``Eval 2" were created to reflect the support analyst working on the support ticket. All of the Support Analyst Profile features mimic the features under the Customer Profile, except that they focus on a single support analyst instead of a single customer.

\subsection{Statistical Results \& Validation - Complete and Final Set of Features}
\label{ssec:stats_final}

\begin{figure}[t]
    \centering
    \includegraphics[width=\columnwidth]{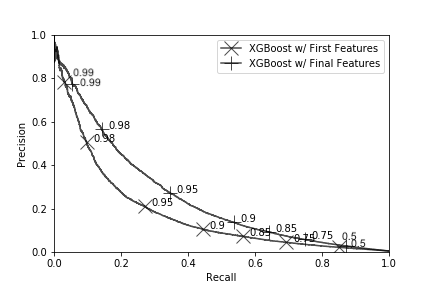}
    \caption{First vs Final Model Results in PR Space (marked with confidence thresholds)}
    \label{fig:prs_original_vs_final_xgb}
\end{figure}

Having engineered all additional features that we identified through the evaluation cycles in our design science methodology, we conducted another statistical validation of the performance of our ML model including all these features. All PMRs and CritSits were randomly distributed into 10 folds, and then 10-fold leave-one-out cross-validation was performed on the dataset using the XGBoost classifier. The results of the validation can be seen in the confusion matrix in Table \ref{table:conf_matrix_final_xgb}. The recall for ``CritSit –- Yes" is 87.36\%, with a precision of 2.79\%, and 88.23\% summarization. These results are an improvement from the first Support Ticket Model Feature results computed with \textit{Random Forest} which had 79.94\% recall, 1.65\% precision, and 80.77\% summarization. The final results were also an improvement over the first Support Ticket Model Feature results computed with \textit{XGBoost} which had 85.14\% recall, 2.16\% precision, and 85.19\% summarization. Figure \ref{fig:prs_original_vs_final_xgb} shows PR Space comparing the performance of the model with the first set vs. final set of features using XGBoost.

For each of the 57 features used in the XGBoost model, a feature importance is reported in Table \ref{table:final_features_2} (``Days since last contact" was not engineered, therefore 57 and not 58). An interesting observation is that 1/5th (11) of the features account for 4/5ths of the total feature importance to the model. These top 1/5th of the feature importances, listed in bold in Table \ref{table:final_features_2}, account for 80.41\% of the total feature importance. Additionally, there are 19 ``0.00"s listed in the table (1/3rd of the features), which indicates that the model was able to gain no benefit from using those features; those features can be removed and the model will produce the same results.

Five of the top 11 features are from the new features created during the evaluation cycles, pointing to the overall benefit gained from continuing to improve the model in collaboration with IBM. In particular, ``Open CritSit to PMR ratio" is the 3rd most important feature in the model at 9.01\% importance.

\section{Evaluation 4: Comparison to Predictions Without FE (RQ3)}
\label{sec:eval4}

\begin{table*}[t]
\centering
\caption{Support Ticket Model Features with Importance Metric (Top 1/5th Important Features in Bold)}
\label{table:final_features_2}
\resizebox{\textwidth}{!}{
\def\arraystretch{1.2}
\begin{tabular}{|c|l|c|c|c|c|c|l|}
    \hline
        \textbf{Category}
        & \textbf{Feature}
        & \multicolumn{5}{c|}{\textbf{Importance (\%)}}
        & \textbf{Description}
    \\\hline\hline
        \multirow{3.5}{*}{\nlcell[1.2]{c}{Basic\\ Attributes}}
            & Number of entries
            & \multicolumn{5}{c|}{\textbf{21.40}}
            & Number of events/actions on the PMR
        \\\cline{2-8}
            & Days open
            & \multicolumn{5}{c|}{\textbf{22.70}}
            & Days from open to close (or CritSit)
        % \\\cline{2-8}
        %     & Escalation type
        %     & \multicolumn{5}{c|}{ --- }
        %     & CritSit Cause, CritSit Cascade, or None
        \\\cline{2-8}
            & PMR ownership level
            & \multicolumn{5}{c|}{\textbf{4.16}}
            & \nlcell{l}{Level of Support (L0 - L3) that is in charge \\ of the PMR, calculated per entry}
    \\\hline\hline
        \multirow{5}{*}{\nlcell[1.2]{c}{Customer\\ Perception\\ of Process}}
            & \nlcell{l}{Number of support people \\ in contact with customer}
            & \multicolumn{5}{c|}{1.83}
            & \nlcell{l}{Number of support people the\\ customer is communicating with}
        \\\cline{2-8}
            & Number of increases in severity
            & \multicolumn{5}{c|}{\textbf{4.56}}
            & Number of times the Severity increase
        \\\cline{2-8}
            & Number of decreases in severity
            & \multicolumn{5}{c|}{\textbf{2.39}}
            & Number of times the Severity decrease
        \\\cline{2-8}
            & \nlcell{l}{Number of sev4/sev3/sev2 \\ to sev1 transitions}
            & \multicolumn{5}{c|}{0.00}
            & \nlcell{l}{Number of changes in Severity\\ from 4, 3, or 2, straight to 1}
    \\\hline\hline
        \multirow{7}{*}{\nlcell[1.2]{c}{Customer\\ Perception\\ of Time}}
            & Time until first contact
            & \multicolumn{5}{c|}{1.37}
            & \nlcell{l}{Minutes before the customer hears from\\ IBM for the first time on this PMR}
        \\\cline{2-8}
            & Current received response time
            & \multicolumn{5}{c|}{1.79}
            & \nlcell{l}{Average number of minutes of all the \\ analyst response times on this PMR}
        \\\cline{2-8}
            & \nlcell{l}{Difference in current vs historical \\ received response time}
            & \multicolumn{5}{c|}{1.77}
            & \nlcell{l}{(Historical received response time) minus \\ (Current received response time)}
        \\\cline{2-8}
            & Days since last contact
            & \multicolumn{5}{c|}{ --- }
            & Number of days since last contact, calculated per entry
        \\\cline{2-8}
            & \nlcell{l}{Difference in historical sent vs \\ historical received response time}
            & \multicolumn{5}{c|}{0.31}
            & \nlcell{l}{(Historical received response time) minus \\ (Historical sent response time)} 
    \\\hline\hline
                & & \multicolumn{5}{c|}{\textbf{Importance (\%) with}} & \\
                & & \multicolumn{5}{c|}{\textbf{Decay After N Weeks *}}
                & \\\cline{3-7}
                & & $\infty$ & 12 & 24 & 36 & 48 &
    \\\hline\hline
        \multirow{8}{*}{\nlcell[1.0]{c}{Customer\\ Profile}}
            & Number of open PMRs
            & 0.23
            & 0.10
            & 0.16
            & 0.11
            & 0.21
            & Number of PMRs owned by customer that are open 
        \\\cline{2-8}
            & Number of closed PMRs
            & \textbf{2.12}
            & 1.53
            & \textbf{2.99}
            & 0.41
            & 0.41
            & Number of PMRs owned by customer that are closed 
        \\\cline{2-8}
            & Number of open CritSits
            & 0.20
            & 0.00
            & 0.07
            & 0.13
            & 0.01
            & Number of CritSits owned by customer that are open
        \\\cline{2-8}
            & Number of closed CritSits
            & 0.04
            & \textbf{3.95}
            & 1.72
            & 0.69
            & 0.60
            & Number of CritSits owned by customer that are closed
        \\\cline{2-8}
            & Open CritSit to PMR ratio
            & \textbf{9.01}
            & ---
            & ---
            & ---
            & ---
            & (Number of open CritSits) / (Number of open PMRs)
        \\\cline{2-8}
            & Closed CritSit to PMR ratio
            & \textbf{4.19}
            & ---
            & ---
            & ---
            & ---
            & (Number of closed CritSits) / (Number of closed PMRs)
        \\\cline{2-8}
            & \nlcell{l}{Historical received response time}
            & 1.39
            & ---
            & ---
            & ---
            & ---
            & \nlcell{l}{Average of all received response times \\ on PMRs owned by this customer}
    \\\hline\hline
        \multirow{7}{*}{\begin{tabular}[c]{@{}c@{}}Support\\ Analyst\\ Profile\end{tabular}}
            & Number of open PMRs
            & 0.39
            & 0.00
            & 0.00
            & 0.00
            & 0.00
            & Number of PMRs owned by customer that are closed
        \\\cline{2-8}
            & Number of closed PMRs
            & 1.24
            & 0.00
            & 0.00
            & 0.00
            & 0.00
            & Number of PMRs owned by the analyst that are closed
        \\\cline{2-8}
            & Number of open CritSits
            & 0.00
            & 0.00
            & 0.00
            & 0.00
            & 0.00
            & Number of CritSits owned by the analyst that are open
        \\\cline{2-8}
            & Number of closed CritSits
            & 0.03
            & 0.00
            & 0.00
            & 0.00
            & 0.00
            & Number of CritSits owned by the analyst that are closed
        \\\cline{2-8}
            & Open CritSit to PMR ratio
            & 1.02
            & ---
            & ---
            & ---
            & ---
            & (Number of open CritSits) / (Number of open PMRs)
        \\\cline{2-8}
            & Closed CritSit to PMR ratio
            & \textbf{2.95}
            & ---
            & ---
            & ---
            & ---
            & (Number of closed CritSits) / (Number of closed PMRs)
        \\\cline{2-8}
            & Historical sent response time
            & 1.82
            & ---
            & ---
            & ---
            & ---
            & \nlcell{l}{Average of all sent response times \\ on PMRs owned by an analyst}
    \\\hline
        \multicolumn{8}{r}{\nlcell{c}{* Each feature importance number represents a unique feature, totalling 58 features in the whole table}}
\end{tabular}
}
\end{table*}

To answer RQ3, which is aimed at verifying enhanced performance through FE, we implemented a baseline approach that implements the XGBoost algorithm in a model that uses all available customer support ticket data without the use of FE, and report on the comparative results to the model versions (``first" as well as ``final" set of features) in our FE approach.

\subsection{Baseline Implementation}

To implement the baseline, we had to feed the model one row of data. This was a required design decision with our dataset because PMRs are composed of multiple entries per PMR (detailed in Section \ref{ssec:ethno}). Therefore, \textit{the last entry before the PMR CritSit date (for CritSits) or closed date (for non-CritSits)} was chosen as the representative data row for each PMR. With this design decision in place, the features of the baseline model are no longer the engineered features in Table \ref{table:final_features_2}, but rather the features from the raw customer support ticket (PMR) data. There are 95 features available in the raw data, but a majority of those features are identification features or strings that are not categorical, which means they cannot be used in ML algorithms without some form of natural language processing. The number of usable features is 34, and similar to Section \ref{sec:eval3} not all features were important to the model and therefore produced a feature importance of 0, leaving the final set of features utilized from the raw data at 25.

Finally, with the data prepared for ML purposes, it was fed through the exact same process of splitting, training, and testing as the process applied to the first and final features.

\subsection{Baseline Results}

\begin{figure}[t]
    \centering
    \includegraphics[width=\columnwidth]{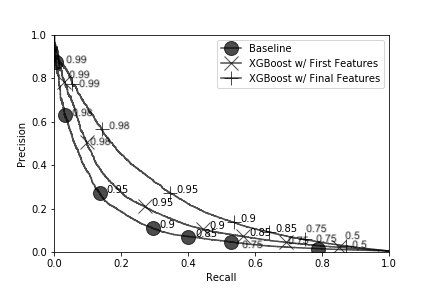}
    \caption{Comparing the performance of the three models using XGBoost: Baseline, First, and Final model features (marked with confidence thresholds)}
    \label{fig:prs_all_graphs}
\end{figure}

The results of the validation can be seen in the confusion matrix in Table \ref{table:conf_matrix_baseline_xgb}. The recall for ``CritSit -- Yes" is 79.04\%, with a precision of 1.54\%, and 80.86\% summarization. These results are only slightly lower than the first features, but are considerably lower than the results obtained when the model included the complete, final set of features. Furthermore, these results are for the chosen threshold of 50\% confidence, a more detailed account of the results are displayed in Fig. \ref{fig:prs_all_graphs} where we show the performance of all three implementations (first set, final set, and baseline) of the models, graphed in PR Space. The baseline implementation has the lowest overall performance, followed by that of the model implementing the first of features, and outperformed by the model implementing the final set of features.

\begin{table}[b]
\centering
\caption{Confusion Matrix for CritSit Prediction using XGBoost on ML model with complete set of Features}
\label{table:conf_matrix_final_xgb}
\resizebox{\columnwidth}{!}{
    \def\arraystretch{1.2}
    \begin{tabular}{|c|c|c|c|}
        \hline
            \multirow{2}{*}{Actual}
            & \multirow{2}{*}{Total}
            & \multicolumn{2}{c|}{Predicted as}
            \\\cline{3-4} 
                & & CritSit - No
                & CritSit - Yes
        \\\hline
            CritSit - No
            & 2,532,745
            & \nlcell{c}{2,242,064 (TN) \\ 88.52\%}
            & \nlcell{c}{290,681 (FP) \\ 11.48\%}
        \\\hline
            CritSit - Yes & 9,536 &
            \nlcell{c}{1,205 (FN) \\ 12.64\%} &
            \nlcell{c}{8,331 (TP) \\ 87.36\%}
       \\\hline
    \end{tabular}
}
\end{table}

\section{Discussion}
\label{sec:discussion}

Prompted by the problem of inefficiency in managing customer support ticket escalations at our industrial partner IBM, our approach had been to study and model the information available to support analysts in assessing whether customers would escalate on a particular problem they reported, and to investigate ML techniques to apply this model to support the escalation management process. We employed a design science methodology and here we discuss, as outlined by Sedlmair et al. \cite{Sedlmair2012}, our contributions through three main design science aspects: problem characterization and abstraction, validated design, and reflection.

\subsection{Problem Characterization and Abstraction}

The investigation of IBM support practices in our ethnographic exploratory study was the first step in our design science iterative process, providing a more detailed understanding of the support ticket escalation problem at IBM. We elaborate here on two lessons learned during the problem characterization phase.

The first lesson we learned is about the importance of this step and iterating through it in the design study. From our initial interviews with the support analysts we were able to draw an understanding of how they work as well as the first set of our PMR model features. However, it was only after the first evaluation step (the in-depth investigation of the ten CritSit PMRs at the Victoria site) that we reflected and refined our understanding of the problem context in the analysts' job. We were able to uncover details of the cascading CritSits process and its effect on how data was being presented to the analysts. This turned out to be crucial to understanding the PMR life-cycle and to refinements in our PMR model features.

\begin{table}[b]
\centering
\caption{Confusion Matrix for CritSit Prediction using XGBoost without FE}
\label{table:conf_matrix_baseline_xgb}
\resizebox{\columnwidth}{!}{
    \def\arraystretch{1.2}
    \begin{tabular}{|c|c|c|c|}
        \hline
            \multirow{2}{*}{Actual}
            & \multirow{2}{*}{Total}
            & \multicolumn{2}{c|}{Predicted as}
            \\\cline{3-4} 
                & & CritSit - No
                & CritSit - Yes
        \\\hline
            CritSit - No
            & 2,557,730
            & \nlcell{c}{2,073,953 (TN) \\ 81.09\%}
            & \nlcell{c}{483,777 (FP) \\ 18.91\%}
        \\\hline
            CritSit - Yes & 9,577 &
            \nlcell{c}{2,007 (FN) \\ 20.96\%} &
            \nlcell{c}{7,570 (TP) \\ 79.04\%}
       \\\hline
    \end{tabular}
}
\end{table}

The second lesson relates to abstracting from the specifics of IBM relative to data that can be modeled for EP in other organizations. We learned that some elements of the support process may be intentionally hidden from customers to simplify the support process for them, but also to protect the organization's information and processes. An example of this is the offline conversations that occur between people working to solve support tickets: a necessary process of information sharing and problem solving, but these conversations are never revealed to customers. Other organizations might have similar practices, and being aware of the distinction between customer-facing and hidden information is important. We recommend that companies experiment with both including and not including information hidden from customers in their ML models. Information not known to their customers may be introducing noise to their models.

\subsection{Validated Support Ticket Model Features}

The two artifacts we iteratively developed in our design science methodology are the Support Ticket Model Features, and their implementation into an EP ML model to assist support analysts in managing sup\-port-ticket escalations. We believe that the major, u\-nique contribution of this research is the Support Ticket Model Features. The features were not only derived from an understanding of support analysts at our industrial partner, but were iteratively refined through several validations of the EP ML techniques that implemented these features.

The task of predicting support-ticket escalations is fundamentally about understanding the customers' experience within the support ticket management process. The features we created in our model were designed to represent the knowledge that support analysts typically have about their customers. Through the process of FE, our work identified the subset of features relevant to EP from an understanding of practice around escalation management. Finally, we sought to abstract from IBM practice towards a general model of the escalation management process, and therefore have our results be applicable to support teams in other organizations.

Once the Support Ticket Model Features had been created, they were used in an EP ML model to investigate the benefit provided to the Support Analysts' job at IBM. Over the course of this research, multiple stages of ML models were created and tested. Among them, the baseline implementation showed the lowest performance results, evident in the PR Space graph. This was expected as the effort that went into this implementation was the lowest, with no FE. The model with the first set of features, created largely from the observations and interviews conducted with IBM, showed improved results over the baseline. Lastly, the model implementing the final and complete set of features produced the best results.

The iterative phases of this research proved important to producing the final results which otherwise may have never been achieved with other methodologies that don't emphasize the feedback cycles present in a design science methodology. The results of the final 10-fold cross-validation (shown in Table \ref{table:conf_matrix_final_xgb}) were the highest of the three implementations, with a recall of 87.36\% and summarization of 88.23\%. Our collaborating IBM support team was very pleased with this result, as an 88.23\% reduction in the workload to identify high-risk PMRs is a promising start to addressing the reduction of CritSits.

Finally, a prototype tool was built to integrate the real-time results of feeding live PMRs data through our model to produce escalation risks. Use of our prototype tool granted shorter meetings addressing more issues focused on support tickets deemed important by IBM and the ML model, while still allowing for longer meetings to review more PMRs if they needed to. The main benefit was the summarization and visualization of the support tickets based on a combination of our model output as well as their own assessment through the MER field.

\subsection{Reflection}

Our work adds to the scarce research into automating the prediction of support ticket escalations in software organizations. We reflect below on the relationship between our work and these existing techniques, and discuss implications for practitioners who wish to use this work.

\subsubsection{Limitations in Comparing to Previous Research}

The work done by both Ling et al. \cite{Ling2005} and Sheng et al. \cite{sheng2014cost} involved improvements to existing cost-sensitive ML algorithms, with no consideration to the features being fed into the model. Our search of the literature makes us classify this work as a non-FE approach. The option of using their work as a baseline to compare precision and recall required our data to be in such a format that it could be run through their algorithms. Our data, however, was not fit for classification-based ML algorithms because it is archival, with multiple historical entries per each support ticket. Basic classification ML algorithms require there to be one entry per support ticket, so any archival data such as ours would have to go through a process to convert that data into a summarized format. The final summarized data depends on the conversion process chosen; therefore, we could not simply convert our data and then hope it conformed to the constraints of the previous studies due to the lack of information regarding their data structures. We could have used the one line approach applied to the baseline, however, the data would have been severely limited in represented the PMR escalation process at IBM. Therefore, comparing with the work of Ling et al. \cite{Ling2005} and Sheng et al. \cite{sheng2014cost} --- as representatives of non-FE approaches --- was not justified given these characteristics of our data. However, in our attempt to compare the performance of our feature-engineered models with that of non-FE approaches, we did implement a baseline that limited the PMR data to a one-row representation as described in Section \ref{sec:eval4}. Analyzing the performance using this representation and our XGBoost algorithm was more rightly justified in light of the PMR escalation process at IBM. 

The work done by Bruckhaus et al. \cite{Bruckhaus2004} has a similar data processing issue, except their work involved some FE to convert their data into a usable form. They neither describe how they conducted their FE nor the final set of engineered features, therefore we could not compare FE results. Furthermore, the details about their neural network approach, including the parameters chosen for their proposed algorithm, are not provided, making its replication difficult.

Given the lack of ability to replicate the process and results of previous work with our data, we were not able to contrast our work against this related work; instead, our research focused on FE and iteratively developing our predictive model with support analysts through our design science methodology.

\subsubsection{New Directions for Further Validating the Features and Model}

Our work represents a first step towards a model of support ticket information through FE relevant to predicting the risk of support ticket escalations; however, further validation of our features and model is needed. In particular, a full evaluation with IBM is needed to address the question of usability and effectiveness inside their organization. This technology transfer project is already underway, and will seek to answer the question of effectiveness inside the organization it was built to help. Additionally, once fully deployed, the Support Ticket Model Features will be further evaluated for both importance to the model as well as importance to IBM in assessing PMRs as potential escalations. The research performed will help shape the final set of features that are used within IBM as a tool for understanding their customers and the escalations that occur.

\subsubsection{Implications for Practitioners}

The model we developed has the potential for deployment in other organizations given that they have enough available data and the ability to map it to the features provided by our model. To implement the ML-based EP model we developed, organizations must track and map their data to the Support Ticket Model Features. If the high recall and summarization we obtained at IBM is obtained at other organizations, there is potential to reduce their escalation identification workload by $\sim$88\%, with the potential for $\sim$88\% of the escalations to remain in the reduced set. If this frees up time for support analysts, then they can put additional effort into more important aspects of the support process like solving difficult issues and identifying bottom-up requirements from support tickets.

Prior to implementing our model, organizations \\should do a cost-benefit analysis to see if the potential benefits are worth the implementation effort. Included in this analysis should be the cost of a support ticket –-- with and without an escalation, as well as time required to manually investigate tickets, customers, and products for escalation patterns. If the overall cost of escalating tickets and the investigative efforts to avoid escalations outweigh the overall time-spent implementing the model described above, then there is a strong case for implementation.

\section{Threats to Validity}

The first threat, to external validity \cite{Shull2008}, is the potential lack of generalizability of the results due to our research being conducted in close collaboration with only one organization. To mitigate this threat, the categories and features in our support ticket model were created with an effort of abstracting away from any specifics to IBM processes, towards data available and customer support processes in other organizations.

The second threat, to construct validity \cite{Shull2008}, applies to the mapping of the information and data we collected through interviews with support analysts to the thematic themes and codes. To mitigate that threat, multiple techniques were used: member checking, triangulation, and prolonged contact with participants \cite{Shull2008}. The design science process of iteratively working with industry through design cycles puts a strong emphasis on member checking, which Lincoln and Guba \cite{lincoln1985naturalistic} describe as ``the most crucial technique for establishing credibility" in a study with industry. We described our themes and codes to the IBM analysts and managers through focus groups and general discussions about our results to validate that our data mappings resonated with their practice. Triangulation, through contacting multiple IBM support analysts at different sites as well as observations of their practice during support meetings, was used to search for convergence from different sources to further validate the features and mappings created \cite{Creswell2000}. Finally, our contact with IBM during this research lasted over a year, facilitating prolonged contact with participants which allowed validation of information and results in different temporal contexts.

The third threat, to internal validity \cite{Shull2008}, relates to the noise in the data discovered during the iterative cycles of our design science methodology. As discussed in Section \ref{sssec:e1_findings}, the CritSits in our dataset could be ``cause" or ``cascade". Due to limitations of our data, we are unable to reliably tell the two types of CritSits apart; however, there is a small subset of CritSits we know for sure are ``cause" CritSits. At the cost of discarding many ``cause" and uncertain CritSits, we removed all ``cascade" CritSit PMRs by discarding the CritSits that had more than one associated PMR. The newer, ``real" CritSit PMRs (CritSits with only one PMR attached) in our data then totaled $\sim$3,500 (35\% of our original target set). The recall on the new target set was 85.38\%, with a summarization of 89.36\%, meaning that the threat to internal validity due to this noise in our data was negligible.

\section{Conclusion}

Effectively managing customer relationships through handling support issues in ongoing software projects is key to an organization's success, and one practice that informs activities of requirements management. Support analysts are a key stakeholder in gathering bottom-up requirements, and proper management of support ticket escalations can allow them to do their job with less attention to escalations.

The data used in this research is confidential, and unfortunately cannot be shared with the research community; furthermore, the algorithms used to transform the data into the engineered features is also confidential, since knowledge of the transformation would give insights into the structure of the data, which is also confidential.

The two artifacts we developed in this work, the Support Ticket Model Features and its implementation in a ML classifier to predict the risk of support ticket escalation, represent a first step towards simplifying support analysts' job and helping organizations manage their customer relationships effectively. We hope that this research leads to future implementations in additional industry settings, and further improvements to EP through ML in future research.

% BibTeX users please use one of
\bibliographystyle{spmpsci}      % mathematics and physical sciences
\bibliography{main}   % name your BibTeX data base

\end{document}